\documentstyle[twoside,fleqn,espcrc2]{article}

\def \be{\begin{displaymath}}
\def \ee{\end{displaymath}}
\def \bea{\begin{eqnarray}}
\newcommand {\vx}       {{\bf x}}
\def \eea{\nonumber \end{eqnarray}}

\begin{document}

\title{Improving  the
Lattice QCD Hamiltonian }

\author{Xiang-Qian Luo, Shuo-Hong Guo\address{ CCAST (World Laboratory),
 P.O. Box 8730, Beijing 100080, 
China\\
Department of Physics, Zhongshan University, 
Guangzhou 510275, China\\ 
Center for Computational Physics, 
Zhongshan University, Guangzhou 510275, China}, 
Helmut Kr\"oger\address{ D\'epartement de Physique, Universit\'e Laval,
Qu\'ebec, Qu\'ebec G1K 7P4, Canada} 
and Dieter Sch\"utte\address{ Institut f\"ur Theoretische Kernphysik,
Universit\"at Bonn, D-53115 Bonn, Germany}
}

\begin{abstract}
\end{abstract}

\maketitle

\section{Introduction and Survey}
Lepage's improvement scheme\cite{Lepage} is a major progress 
in the recent development of lattice QCD,
opening the possibility to approach continuum physics 
on very coarse lattices.
For the purpose of profiting from this scheme
in the Hamiltonian formulation, 
the problem of improving the lattice Hamiltonian
for the pure gauge theory is investigated.
Following the procedure of Lepage,
first a classically improved Kogut Susskind Hamiltonian
is derived using
a transfer matrix\cite{Munster} or Legendre 
transformation\cite{Legendre} method.
In this formulaltion, the color electric energy
becomes an infinite series with long range terms.
This deficiency can be cured by
arguing with the non-uniqueness of the
classically improved action yielding the
result that the same order of improvement
can be achieved by keeping only
nearest neighbor interactions.
A further tadpole improvement\cite{Lepage} of the
Hamiltonian may be introduced straightforwardly.
A final improvement in the sense of L\"uscher and
Weiss\cite{Luscher} involves a (Hamiltonian) perturbative
 calculation of suitable observables
which will be deferred to the future.
Here we will only discuss the qualitave
structure of the occuring new Hamiltonian terms.
An incorporation of the improved Hamiltonian
into a coupled cluster calculation\cite{Luo,Sch} of the
spectrum is possible and should allow to obtain
reliable results for the glueball spectrum
in lower order than with the original
Kogut Susskind Hamiltonian.

\section{ Hamiltonian Formalism}
Starting from a lattice action written in terms
of link variables $U_{x,\mu}$, the related Hamiltonian 
may be constructed by the following procedure.
First one has to differentiate  spacelike ($a$) and 
timelike ($a_0$) lattice spacings, and one has to 
introduce a temporal gauge: $U_{x,0} = 1$.
The action may then be decomposed in the form
(for details see \cite{Munster})
\be 
S = \frac{a}{a_0}S^0 + \frac{a_0}{a} S^1\;,
\nonumber
\ee
where $S^1$ {\em does} not couple different times
\be
S^1 =\sum_t S^1(U_t)\; .
\nonumber
\ee
Here, $U_t$ stands for the {\em set} of (spacelike)
link variables at fixed time $t$,
$U_t = \{U_{\vx,t,j}\}$.
For small $a_0$, 
$S^0$ may be restricted to have the structure 
\be
S^0 =\sum_t  S^0(U_t,V_t, V^\dagger_t)\;,
\ee
where
$V_t = \{U_{\vx,t+a_0,j} U^\dagger_{\vx,t,j}\}$.
For $a_0 \rightarrow 0$ the elements of the set $V_t$
are close to unity
and may be characterized by the Lie algebra of $SU(N_c)$
\be
U_{\vx,t+a_0,j} U^\dagger_{\vx,t,j} = 1 + i q_{\vx,j}(t) 
-\frac12 (q_{\vx,j}(t))^2
\nonumber
\ee
where $q_{\vx,j} (t) = \sum_b q_{\vx,j,b}(t) T^b$. The general
structure of $S^0$ is then
\bea
S^0 &=& \sum_t \sum_{\alpha,\beta} \frac12 
q_{\alpha}(t) M_{\alpha,\beta}(U_t) q_\beta(t)\;,
\nonumber \\
\alpha &=& (\vx,j,b)\;.
\end{eqnarray}

The  {\em Hamiltonian} related to the 
action $S$ is then defined on 
Hilbertspace functions
$\psi(U)$ ($t$ fixed) and is given by
\begin{eqnarray}
H &=& T + V
\nonumber \\
T &=&
\frac{1}{2a}  \sum_{\alpha,\beta} 
E_{\alpha} M^{-1}_{\alpha,\beta}(U) E_\beta
\\ \nonumber
V &=&
\frac1a S^1(U) \;,
\end{eqnarray}
where $E_{\vx,j} = E_{\vx,j,b} T^b$ is conjugate to $U_{\vx,j}$.
The validity of these formulae can be derived using the
transfer matrix\cite{Munster} or the
Legendre transformation\cite{Legendre} method.
\\

The simplest example for the action $S$ is the  Wilson action
$S_{Wilson} = \beta \sum_{x, \mu>\nu} (1-P_{\mu\nu}(x))$
 ($P_{\mu\nu}$ is the standard plaquette parallel transporter.) 
yielding
$M_{\alpha,\beta} = g^{-2} \delta_{\alpha,\beta}$
and the Kogut-Susskind Hamiltonian 
\be
H_{KS} = \frac{g^2}{2a}  \sum_{\alpha} 
E_{\alpha} E_\alpha
+ \frac{2N_c}{ag^2} \sum_{\vx, j>k} (1-P_{jk}(\vx))
\nonumber
\ee

\section{The Classically Improved Hamiltonian}
The classiscal $O(a^2)$ errors of the Wilson action
are compensated by introducing the improved action\cite{Lepage}
(we disregard constant terms)
\be
S_{cl}=-\beta \sum_{x, \mu>\nu} 
\left[\frac53 P_{\mu\nu}(x))-\frac{1}{12}
(R_{\mu\nu}(x)+R_{\nu\mu}(x))\right] 
\ee
where $R_{\mu\nu}$ is 
the $2a \times a$ "rectangle loop parallel transporter". 

Differentiating space- and time-like lattice spacings the
improved potential part of the Hamiltonian
is given by the space part of $S_{cl}$
\be 
V_{cl}= -\frac{\beta}{a}\sum_{\vx, j>k}\left(\frac53
  P_{jk}(\vx)
 - \frac{1}{12}(R_{jk}(\vx)+R_{kj}(\vx))\right)
\nonumber
\ee
The kinetic part involves a non-trivial computation:
The $S^0$-part of $S_{cl}$ contains
the time-like terms $P_{i0}$ and $R_{i0}$ 
which have  
up to errors of the order $O(a^4,a_0^2)$
the continuum limit behaviour
(in the limit $a_0 \rightarrow 0$ errors of $O(a^4_0)$
can be disregarded)
\bea
P_{i0}&=&-\frac{a_0^2 a^2}{\beta}
\left(  TrF_{i0}^2 
+ \frac{a^2}{12} Tr F_{i0}(D_j^2)F_{i0} \right)
\nonumber \\
R_{i0}&=&-\frac{a_0^2 a^2}{\beta} \left( 4TrF_{i0}^2 
+ \frac{16a^2}{12} Tr F_{i0}(D_j^2)F_{i0} \right)
\eea
Consequently, a classical improvement of Wilson's $S^0$
is given by
\be 
S^0_{cl} = 
 -\beta \frac{4}{3}\left(
 \sum_{x, j} P_{0j}(x)
 - \frac{1}{16}\sum_{x, j} R_{0j}(x)\right)
\nonumber
\ee

With temporal gauge fixing and expanding 
according to (1) this yields for the matix M
\bea
&&M_{\alpha,\alpha'} = \frac{4g^2}{5} \left[\delta_{\alpha,\alpha'}
\right. \nonumber \\
&&\left. - \frac{1}{20} \delta(\vx',\vx+j)\delta_{j,j'}
Tr \;U^\dagger_{\vx,j} \;T^a \;U_{\vx,j} \;T^{a'}\right]
\nonumber
\eea
The inversion of the matrix M yields for the 
kinetic part $T$ of the classically
improved Hamiltonian  
an infinite number of terms
\bea
M^{-1}_{\alpha,\alpha'} = \frac{5}{4g^2} \left[\delta_{\alpha,\alpha'}
+\;\;\;\;\;\hspace{4cm} \right.&& \nonumber\\ \nonumber\left.
\sum^\infty_{n=1}(\frac{1}{20})^n \delta(\vx',\vx+nj)\delta_{j,j'}
Tr \;U^\dagger_{\vx,nj} \;T^a \;U_{\vx,nj} \;T^{a'}\right]&&
\eea
where $U_{\vx,nj}$ is the straight line parallel transporter from
$\vx$ to $\vx + nj$, \\
$U_{\vx,nj} = U_{\vx,j}U_{\vx+j,j}...U_{\vx+(n-1)j,j}$.
Note that the combination of the parallel transporters
$U_{\vx,nj}$  and the 
conjugate operators $E_\alpha$ 
in (2) guarantees the gauge invariance
of $T$.

\section{Classically Improved Hamiltonian
with Finite Number of Terms}
The occurance of infinite number of terms in $T$
can be avoided by invocing the fact that the classically improved action 
given by demanding cancellation of errors up 
to the order $O(a^4)$ is not  uniquely fixed.
Herefore we note that the term
\bea
S_{nj} &=& \sum_{\vx,a} q_{\vx,j,a}\delta(\vx,\vx+nj)\delta_{j,j'}
\\ \nonumber
& & \times Tr \;U^\dagger_{\vx,nj} \;T^a \;U_{\vx,nj} \;T^{a'}
q_{\vx+nj,j',a'}
\eea 
emerges from a generalization of the $2a \times a_0$
loop term $R_{j0}$
to a $na \times a_0$ loop parallel transporter
 $R_{nj,0}$ of the type

\setlength{\unitlength}{0.012in}%
\be
R_{nj,0} = \frac{1}{N_c} Re\;Tr\ \ 
\begin{picture}(130,0)(25,765)
\thinlines
\put( 25,785){\line( 0,-1){ 30}}
\put( 25,755){\line( 1, 0){ 130}}
\put(155,755){\line( 0, 1){ 30}}
\put(155,785){\line(-1, 0){ 30}}
\put(125,785){\line( 0,-1){ 25}}
\put(125,760){\line(-1, 0){ 70}}
\put( 55,760){\line( 0, 1){ 25}}
\put( 55,785){\line(-1, 0){ 30}}
\end{picture}
\nonumber
\ee
 in the sense that we have with temporal
gauge fixing and for small $a_0$
\be
1-R_{nj,0} =\frac{a}{\beta}\left(\sum_{\vx,a} q_{\vx,j,a}
q_{\vx,j,a} + S_{nj}\right)
\ee
The continuum limit behaviour is in this
case
\bea
R_{nj,0}&=&-\frac{ a_0^2 a^2}{\beta}\left[ 4 TrF_{0j}^2 \right.
\nonumber \\ \nonumber 
&+& \left.\frac{(3n^2+1)a^2}{3} Tr F_{0j}(D_j^2)F_{0j}
+ O(a^4,a_0^2)\right]
\eea
Consequently, the ansatz
\be 
S^0=-\frac{\beta}{A}\left(B \sum_{x, j} P_{0j}(x)
 +\sum_{n=1}^\infty C^n
 \sum_{x, j} R_{nj,0}(x)\right)
\ee
is legitimate for a consistent continuum limit, 
and yields  with $B=1-\sum_{n=1}^\infty C^n$
for $M$ {\em a geometrical} series
\bea
M_{\alpha,\alpha'}=g^{-2} A^{-1} \left[\delta_{\alpha,\alpha'}
\;\;\;\;\;\hspace{3cm}\right. &&\nonumber\\ \nonumber \left.
+ \sum_n C^n \delta(\vx',\vx+nj)\delta_{j,j'}
Tr \;U^\dagger_{\vx,nj} \;T^a \;U_{\vx,nj} \;T^{a'}\right]&&
\eea
Hence $M^{-1}$ has only two terms
\bea
M^{-1}_{\vx j a;\vx'j'a'}=g^2 A \left[\delta_{\vx j a;\vx'j'a'}
\hspace{3cm}\right.&&\\ \nonumber\left.
-C \delta(\vx',\vx+j)\delta_{j,j'}
Tr \;U^\dagger_{\vx,j} \;T^a \;U_{\vx,j} \;T^{a'}\right]&&
\eea
A classical  improvement of $T$ to the order $O(a^4)$ is
then obtained for
$A=(1+c^2)/(1-c^2)$ and $ C = 2c/(1+c^2)$
where $c\simeq -.101$ is a solution of
$\sum_{n=0}^\infty (1 + 6n^2) c^n = .5$.

\section{Further Improvements}
The {\em tadpole improvement} may be defined as
in the Lagrangian case\cite{Lepage} by replacing
$U_{\vx,j}$ by $U_{\vx,j}/u_0$
where the quantity $u_0$ may be computed
from the vacuum expectation  value of the
plaquette operator
via
$u_0 = <0|P_{jk}|0>^{1/4}$.
For a L\"uscher Weisz improvement\cite{Luscher}, a
new perturbative, Hamiltonian calculation
of suitable observables has to be performed
which has still to be done.
We expect that the type of additional terms
occurring in the Hamiltonian formulation
correspond to those of Lepage.
The potential part of the improved
Hamiltonian would then get corrections like
the action of Lepage (see eq. (73) of ref.\cite{Lepage}).
It might be interesting to note, however,
that the 
 time-like ``parallelogram plaquettes''
(eq. (72) of ref.\cite{Lepage})
yield contributions to $T$ which are 
quadratic and linear  in the opertors
$E_\alpha$ combined with non-straight parallel-transporters
(plaquettes in lowest order). Details wil be presented
elsewhere\cite{Kroger}.

XQL and SHG were supported by the National Natural Science 
Foundation and National Education Committee.
H.K. was supported by NSERC Canada.

\end{document}